\documentclass[twocolumn,floatfix]{revtex4-2}

\usepackage{amsmath,amssymb,amsfonts}
\usepackage{graphicx}
\usepackage{physics} 
\usepackage[ruled,vlined,linesnumbered]{algorithm2e}
\SetKwInOut{KwIn}{Input}
\SetKwInOut{KwOut}{Output}

\usepackage{adjustbox}
\usepackage{nicematrix}
\usepackage{placeins}
\usepackage{tikz}
\usepackage{dsfont}
\usepackage{makecell} 

\usepackage[colorlinks=true,linkcolor=blue,citecolor=blue,urlcolor=blue]{hyperref}

\newcommand{\dashedmid}{
  \mathrel{\kern0.15em
    \tikz[baseline={(0,0)}] \draw[dashed] (0,0) -- (0,1.2em);
  \kern0.2em}
}
\newcommand{\id}{\mathds{1}}

\newcommand{\dA}{\widetilde{\delta}_A}
\newcommand{\dB}{\widetilde{\delta}_B}
\newcommand{\dC}{\widetilde{\delta}_C}
\newcommand{\dD}{\widetilde{\delta}_D}

\begin{document}

\title{Single-Shot Decoding of Biased-Tailored Quantum LDPC Codes}

\author{Devon Campbell}
\email{dec2180@columbia.edu}
\affiliation{Columbia University}

\date{\today}

\begin{abstract}
Quantum processors are often affected by \emph{biased noise} and \emph{noisy readout}, which reduce reliability and reproducibility. This work combines two complementary strategies to address these challenges. The first is \emph{bias tailoring}, which aligns stabilizers with the dominant error type. The second is \emph{single-shot (SS) decoding}, which uses metachecks to identify measurement faults from just one noisy round. We implement these ideas in a four-dimensional lifted hypergraph-product (4D-LHP) code constructed from quasi-cyclic protograph seeds. Simulation results show that bias tailoring lowers the word-error rate (WER) by 20--60\% across realistic $Z{:}X$ bias ratios (from $1{:}1$ up to $1000{:}1$), with the largest improvements at moderate bias. When measurement noise is present, a single SS round recovers more than one third of the performance lost to readout errors. Moreover, metachecks identify over $99.8\%$ of faulty syndromes, providing near-complete fault visibility even with limited correction power. Together, these findings demonstrate that 4D-LHP codes maintain strong resilience under realistic noise, making them promising candidates for integration into orchestrated QPU--CPU workflows.
\end{abstract}

\maketitle

\pagestyle{plain} 

\section{Introduction}
\vspace{-0.5em}
Noisy intermediate-scale quantum (NISQ) devices are fundamentally constrained by the high error rates of quantum bits (qubits), making robust error correction essential for reliable computation \cite{Preskill2018-zv}. However, quantum error correction (QEC) is itself an imperfect process, which can misidentify errors or introduce new ones through faulty corrections \cite{Shor1996-ct}. To mitigate the introduction of errors during QEC, stabilizer measurements are performed repeatedly and analyzed via classical post-processing to estimate the most likely syndrome outcomes \cite{Bombin2015-sf}. 

However, the number of required repetitions typically scales with code size, increasing circuit depth, execution time, and the number of physical operations—all of which exacerbate the likelihood of introducing noise \cite{Bombin2015-sf}.
Alternative strategies have been proposed to improve the reliability of stabilizer measurements. For instance, \textit{single-shot} (SS) error correction avoids repeated measurements by enabling the correction of both data and measurement errors in a single noisy measurement round, significantly reducing QEC overhead \cite{Bombin2016-oe}. 

SS decoding relies on geometric redundancy: additional structure in the code that enables \emph{metachecks}, which apply parity constraints directly to the stabilizer measurement outcomes \cite{Bombin2015-sf}. This redundancy allows a single noisy syndrome extraction to reveal information about both the underlying data error and any faults in the measurement process itself.

Conventional stabilizer codes are typically two-dimensional, with independent $X$- and $Z$-type checks. However, 2D constructions lack sufficient degrees of freedom to impose independent parity checks on the syndrome bits. To support metachecks, one must increase the dimensionality of the code to embed additional constraints on the measured syndromes.

For example, \emph{three-dimensional} homological product codes achieve SS decoding in the \emph{infinite-bias limit}, where only a single Pauli error type (e.g., $Z$) is dominant. In this regime, the code can detect and correct both data and measurement errors, but it only supports metachecks for the relevant stabilizer type—leaving the complementary sector (e.g., $X$ stabilizer checks) unprotected \cite{Quintavalle2021-hr}.

To achieve full single-shot capability under mixed Pauli noise, an additional homological layer is required. Four-dimensional chain complexes provide two distinct metacheck matrices, $M_X$ and $M_Z$, which impose independent constraints on $X$- and $Z$-type stabilizers, respectively \cite{Campbell2019-ez}. This structure enables the simultaneous detection of measurement faults in both sectors, supporting reliable single-round decoding under realistic, circuit-level noise.

Even when temporal overhead is minimized, QEC still demands many physical qubits to realize a single logical qubit. Most canonical codes are engineered for \emph{depolarizing} noise, where bit-flip ($X$) and phase-flip ($Z$) errors occur with equal probability. In real hardware, however, the error channel is often \emph{biased}: one Pauli error dominates while the others are strongly suppressed. For instance, superconducting cat qubits exponentially suppress bit-flips, producing a noise profile in which $Z$ errors vastly outnumber $X$ errors \cite{Ruiz2025-hx}.

Ignoring this asymmetry squanders coding resources. A depolarizing-optimized code allocates half of its check operators to detect the minority error, leaving the dominant error under-protected. The result is a lower effective distance in the high-bias direction, a dramatically reduced threshold, and a larger qubit overhead to reach a given logical-error target. In extreme cases, the logical error rate can be \emph{worse} than if one simply encoded the qubit in a repetition code matched to the dominant error. Consequently, a code that remains “un-twisted’’ in a biased channel wastes parity checks on rare events while offering little resistance to the common ones.

To exploit the underlying asymmetry, researchers have devised bias-preserving gates and \emph{bias-tailored} codes that realign stabilizers with the physical noise, cutting overhead and boosting performance \cite{Puri2020-vm}. A prominent example is the \textit{XZZX surface code}, obtained by applying a Hadamard rotation to half of the qubits so that each plaquette involves a balanced mix of $X$ and $Z$ operators \cite{Bonilla_Ataides2021-id}. 

In the infinite-bias limit this “twist’’ decomposes the lattice into independent repetition codes; the logical $Z$ distance grows, allowing the code to tolerate $Z$-error rates up to 50 \%. Even at moderate bias the threshold climbs to $=10.9\%$, compared with $1\%$ for the untailored surface code.

Geometrically, the twist forces logical operators—originally short, straight-line paths across the untwisted surface code—to follow longer, diagonal trajectories. This increases the number of dominant errors required to span the code and cause a logical failure. In short, bias tailoring converts hardware asymmetry from a liability into an advantage, whereas neglecting bias leaves the code fragile along its most vulnerable axis.

Further, bias-tailoring is not confined to the XZZX surface code. The XZZX construction can be understood as a special instance of the \emph{hypergraph product} (HGP) code, a broader framework introduced by Tillich and Zémor \cite{Tillich2014-cj}. This construction takes as input two classical binary linear codes—specified by their parity-check matrices \(H_1\) and \(H_2\)—and produces a quantum CSS code whose stabilizers are derived from the tensor product structure of the seeds. The resulting quantum code inherits the LDPC properties of the classical inputs and can achieve favorable distance and rate scaling depending on the seed choices.

In particular, the XZZX surface code arises when both classical seeds are length-\(L\) repetition codes. This yields a square lattice with stabilizers formed from combinations of \(X\) and \(Z\) operators on four-qubit plaquettes. A local Hadamard rotation on half the qubits then transforms the original surface code into its XZZX variant, aligning the stabilizers to better withstand biased noise. 

More generally, Roffe \textit{et al.} showed that the same “twist’’—a tailored Hadamard rotation that reorients the stabilizer structure—can be applied to \emph{any} HGP code \cite{Roffe2023-ay}. Their work then investigates the \emph{lifted product} framework introduced by Panteleev and Kalachev, which generates sparse quantum codes from a pair of quasi-cyclic matrices and achieves asymptotically growing distance and rate \cite{Panteleev2022-bp}. 

By adapting the twisting procedure beyond repetition codes to arbitrary seed pairs, Roffe’s approach broadens the applicability of bias tailoring from a single surface-code variant to an entire class of quantum LDPC codes. This unlocks new opportunities for optimizing rate–distance trade-offs in the presence of asymmetric noise.

This work combines bias-tailoring design principles with the \emph{single-shot} (SS) decoding paradigm. Starting from \emph{four} classical seeds \(\{\delta_A, \delta_B, \delta_C, \delta_D\}\), we build a four-dimensional lifted hypergraph-product (4D-LHP) code.

Two key features distinguish this construction:

\begin{itemize}
\item \textbf{Hadamard-based bias tailoring.} By applying local Hadamard rotations to half of the qubits, the code’s stabilizers are “twisted” in the manner of bias-tailored lifted product codes. This operation realigns the stabilizer support to increase the minimum weight of logical operators in the dominant error basis—effectively increasing the distance of the code under dephasing-biased noise.
\item \textbf{Metachecks for single-shot decoding.} Additional parity checks, known as metachecks, are embedded one homological layer above and below the conventional stabilizer layers. These checks enable single-shot decoding à la Bombín \cite{Bombin2015-sf,Campbell2019-ez}, allowing both data and measurement faults to be inferred from a single round of noisy syndrome extraction.
\end{itemize}

These mechanisms address complementary challenges: bias tailoring improves resilience against \emph{data errors}—particularly those concentrated in a single Pauli basis—while SS decoding mitigates \emph{measurement errors}, which would otherwise necessitate repeated and costly rounds of syndrome readout. Their combination results in a doubly protected architecture with minimal overhead.

The remainder of this paper is organized as follows. Section~\ref{sec:background} reviews the relevant background on stabilizer codes, biased noise, and single-shot decoding. Section~\ref{sec:methods} presents the 4D-LHP construction, detailing the chain complex, metacheck structure, and bias-tailored Hadamard transformation. Further, this section describes the simulation methodology, including the noise model, decoding algorithm, and protograph lifting. Section~\ref{sec:results} reports numerical results, evaluating performance under varying bias and measurement noise. Finally, Section~\ref{sec:disc} discusses the synergy of the two design principles, potential optimizations, and directions for future work.

\vspace{-1em}
\section{Background}
\label{sec:background}
\vspace{-0.5em}

\subsection{Stabilizer Codes}
\vspace{-0.5em}

A quantum error-correcting (QEC) code typically uses \(n\) physical qubits to encode \(k\) logical qubits. The resulting code is denoted as \([n, k, d]\), where \(d\) is the \emph{distance}—the minimum number of physical qubit errors required to induce a logical failure. A larger distance implies stronger error protection. 

Error correction is achieved by defining a set of mutually commuting operators called \textit{stabilizers}, which constrain the system to a protected subspace known as the code space. A stabilizer code has \(n-k\) independent stabilizer generators \(S_i\), each composed of Pauli operators acting on the \(n\) qubits. Valid codewords are quantum states \(\ket{\psi}\) that satisfy \(S_i \ket{\psi} = \ket{\psi}\) for all \(i\). 

When errors occur, measuring the stabilizers yields a binary syndrome that encodes partial information about the type and location of the error. The decoder uses this syndrome to infer a correction that returns the system to the code space, ideally without disturbing the encoded logical state.

A CSS code is a stabilizer code defined by two binary parity-check matrices $H_X$ and $H_Z$, which determine the structure of the stabilizer generators:
\begin{enumerate}
    \item Each row of $H_X$ defines an \textbf{$X$-type stabilizer} (composed of Pauli-$X$ operators), which detect $Z$ errors (phase-flips) \vspace{-0.5em}
    \item Each row of $H_Z$ defines a \textbf{$Z$-type stabilizer} (composed of Pauli-$Z$ operators), which  detect $X$ errors (bit-flips)
\end{enumerate}

These matrices must satisfy the commutation condition: $H_XH_Z^T = 0 \in \mathbb{F}_2$. Further, the columns of these matrices correspond to the physical qubits that upon which these stabilizers act. For any entry \( a = (i, j) \in H_X \), a value of \( a = 1 \) indicates that stabilizer \( S_i \) acts nontrivially on qubit \( j \), and can detect a \( Z \)-error via anticommutation. Similarly, entries in \( H_Z \) indicate sensitivity to \( X \)-errors.

Suppose an error $e \in \mathbb{F}^n_2$ occurs on the physical qubits. The syndromes associated with that error are: 
$s_X = H_Xe,\quad  s_Z = H_Ze$. If $s_X \neq 0$, at least one $X$=type stabilizer detected an inconsistency, implying there was likely a $Z$-type error. The same logic applies to $s_Z$ for $X$-errors. The role of the \textit{decoder} is to infer a correction operation $r \in  \mathbb{F}^n_2$ based on the observed syndrome $s$, such that applying $r$ cancels out the error and returns the system to the code space. 

The hypergraph product (HGP) constructs a CSS code from any pair of classical binary codes with parity-check matrices $H_1 \in \mathbb{F}_2^{m_A \times n_A}$ and $H_2 \in \mathbb{F}_2^{m_B \times n_B}$. The stabilizer generators are given by:  

\begin{align}
\label{hgp}
H_{\mathrm{HGP}} &= [H_X \;|\; H_Z]  \\
&= \left[
\begin{array}{cc|cc}
0 & 0 & \mathbb{I}_{n_1} \otimes H_2 & H_1^T \otimes \mathbb{I}_{m_2} \\
H_1 \otimes \mathbb{I}_{n_2} & \mathbb{I}_{m_1} \otimes H_2^T & 0 & 0
\end{array}
\right] \notag
\end{align}

This guarantees $H_X H_Z^T = 0$, so the stabilizers commute and define a valid CSS code. The resulting  code is sparse, LDPC, and inherits its logical structure and distance scaling from the input codes \cite{Tillich2014-cj}.

More generally, the HGP construction is a special case of the \emph{homological product code} framework. In this setting, stabilizer codes are modeled using a chain complex—a sequence of vector spaces \(\mathcal{C}_j\) connected by boundary maps \(\delta_j\) satisfying \(\delta_j \circ \delta_{j+1} = 0\) \cite{homological, Bravyi2013-gu}. Each space \(\mathcal{C}_j\) represents a collection of objects (e.g., checks or qubits), and the maps encode how those objects interact. 

The \emph{dimension} of a homological code refers to the length of this chain complex. For example, a two-dimensional code, like those produced by the HGP construction, corresponds to a chain complex consisting of three vector spaces connected by two boundary maps,
: 
\[
\mathcal{C}_{-1} \xrightarrow{\delta_{-1}} \mathcal{C}_0 \xrightarrow{\delta_0} \mathcal{C}_1,
\]
\noindent
where $\mathcal{C}_0$ represents the physical qubits, $\mathcal{C}_1$ represents the $Z$-type stabilizer checks, and $\mathcal{C}_{-1}$ represents the $X$-type stabilizer checks. The maps $\delta_0$ and $\delta_{-1}$ correspond to the parity-check matrices $H_X$ and $H_Z^T$, respectively. 

This structure allows one to generalize the construction to higher dimensions by adding additional layers of checks and metachecks, as is required for single-shot decoding \cite{Bombin2015-sf}. In particular, increasing the number of classical seed matrices used in the tensor construction extends the length of the chain complex: each additional seed introduces a new homological layer and a corresponding boundary map. 

For example, while the HGP uses two classical codes and forms a length-2 complex, a 3D homological product code uses three seeds to form a length-3 complex \cite{Quintavalle2021-hr}, and a 4D construction requires four seeds and yields a length-4 chain. This expansion enables the inclusion of metachecks—higher-layer parity checks that constrain the syndromes themselves—and is essential for achieving full single-shot decoding in the presence of general Pauli noise.

\vspace{-0.5em}
\subsection{Single-Shot Decoding}
\vspace{-0.5em}
In realistic devices, syndrome extraction is itself error-prone: each stabilizer measurement may flip due to gate or readout noise. The conventional solution is to repeat the measurement several times and use temporal correlations to infer the true syndrome \cite{Campbell2019-ez}.

By contrast, single-shot (SS) codes can tolerate a single round of noisy syndrome measurements and still reliably identify data and measurement errors. To achieve this, SS codes are designed with redundant stabilizers that ensure measurement faults imprint a consistent and detectable signature—known as the \emph{metasyndrome}—onto the measured syndrome \cite{Campbell2019-ez}.
  
This redundancy is encoded through metachecks, which impose consistency conditions on syndrome measurements. Formally, the metacheck matrix \(M\) imposes a consistency condition on the observed syndrome \(s\). For a physical error \(E\), the syndrome map \(\sigma(E)\) satisfies \(Ms = M\sigma(E) = 0\) in the absence of measurement noise. If a measurement error \(u\) occurs, the observed syndrome becomes \(s = \sigma(E) + u\), and the metacheck output becomes \(Ms = Mu\). Because the metacheck output depends only on the measurement error $u$, a nonzero metasyndrome $Ms$ signals the presence of measurement faults and can be used to localize and correct them.

Recent work proved that \emph{three-dimensional} homological product (3D-HP) codes exhibit \textit{confinement}, a structural property that plays a critical role in enabling single-shot decoding \cite{Quintavalle2021-hr}. Confinement ensures that the effects of physical and measurement errors remain spatially localized, preventing small faults from producing widespread, correlated syndrome patterns. This locality makes the decoding problem tractable even when syndromes are noisy: faulty measurements generate constrained, predictable patterns that can be detected and corrected without requiring repeated rounds of syndrome extraction.

In the 3D setting, confinement is sufficient to support single-shot decoding in the \emph{infinite-bias} limit, where only one Pauli error (e.g., $Z$) dominates. However, because 3D-HP codes can only embed metachecks for a single stabilizer type, they are not fully single-shot under general Pauli noise. 

From a practical standpoint, these codes can be decoded efficiently using \emph{belief propagation and ordered statistics decoding} (BP+OSD), a two-stage strategy that combines speed with robustness. The first stage applies BP to estimate a likely error configuration by passing probabilistic messages across the Tanner graph. 

The second stage then refines this estimate using OSD, a greedy search that selects low-weight corrections consistent with the observed syndrome \cite{Panteleev2021-oz}. This approach handles both data and measurement noise and is well suited to codes with sparse, redundant checks—conditions naturally met by homological product constructions.

Importantly, BP+OSD benefits from \emph{syndrome redundancy}—the presence of overlapping constraints that provide multiple ways to validate or refute a syndrome bit. In HP codes, this redundancy arises geometrically: each physical qubit or stabilizer participates in multiple tensor-product layers, allowing small faults to be detected through several independent syndromes. This multi-layered structure not only boosts detection but also improves decoder convergence and accuracy under realistic noise.

In the 4D-LHP code introduced in this work, these principles are extended one dimension higher. The use of four classical seed matrices generates a length-4 chain complex, in which both $X$- and $Z$-type stabilizers are independently metachecked. The same BP+OSD machinery applies to each decoding stage: one round to clean the noisy syndrome using metachecks, and another to recover from data errors using the corrected syndrome. Together, these steps constitute a practical, scalable single-shot decoder for general Pauli noise.

\vspace{-1em}
\subsection{Exploiting Bias via Stabilizer Design}
\vspace{-0.5em}

\subsubsection*{XZZX Surface Code}


Exploiting biased noise has become a prominent strategy to improve QEC performance. The archetypal example of this strategy is the \textit{XZZX surface code}: by applying a Hadamard to half of the qubits on a square lattice, the code's stabilizers change from pure-$X$ and pure-$Z$ plaquettes, to mixed four-body XZZX operators 
\cite{Xu2023-fw, Bonilla_Ataides2021-id}. This Hadamard rotation has two immediate impacts under a dephasing-dominated channel ($p_Z \gg p_X$)
\begin{enumerate}
    \item \textbf{Distance boost.} Logical $Z$ operators must follow longer, diagonal paths across the lattice, increasing the minimum weight for damaging $Z$-error chains. For a rotated $L \times L$ code, this boosts the $Z$-distance from $O(L)$ to $O(L^2)$ in ideal bias-tailored layouts. More generally, the logical $Z$-distance of an $N$-dimensional Clifford-deformed code scales as $O(L^N)$ \cite{Huang2023-ni}.
    \item \textbf{Higher thresholds.} In the infinite bias regime, where only $Z$ errors occur, the XZZX code achieve a threshold of $p_Z^{\mathrm{th}}=50\%$ \cite{Bonilla_Ataides2021-id}. Under moderate bias, such as a 4:1 ratio of $Z$ to $X$ errors, the threshold remains high at $p_Z^{\mathrm{th}} \approx 10\%$. Both biased thresholds represent substantial improvements over the threshold of the standard surface code under depolarized noise: $p^{\mathrm{th}} \approx 1\%$ \cite{Fowler2009-xe}.
\end{enumerate}

Roffe \textit{et al.} extended the bias-tailoring strategy of the XZZX code to general quantum LDPC constructions by incorporating Hadamard rotations directly into the code’s structure. Since the columns of the parity-check matrices \( H_X \) and \( H_Z \) index physical qubits, applying Hadamard gates to half of the qubits corresponds to applying the Hadamard transformation to the matching columns in both matrices. 

Because the Hadamard swaps Pauli bases (\( X \overset{H}{\leftrightarrow} Z \)), this operation effectively exchanges the roles of \( X \)- and \( Z \)-type checks on the targeted qubits. As a result, applying Hadamards to half of the qubits is equivalent to swapping the rightmost halves of \( H_X \) and \( H_Z \) in Eq.~\ref{hgp}, yielding mixed-type stabilizers optimized for dephasing-biased noise (Eq.~\ref{hgp_swap}).

\begin{align}
\label{hgp_swap}
H' &= \left[ H_{X_1} \ H_{Z_2} \;\middle|\; H_{Z_1} \ H_{X_2} \right] \notag \\
&= \left[
\begin{array}{cc|cc}
0 & H_1^T \otimes \mathbb{I}_{m_2} & \mathbb{I}_{n_1} \otimes H_2 & 0 \\
H_1 \otimes \mathbb{I}_{n_2} & 0 & 0 & \mathbb{I}_{m_1} \otimes H_2^T
\end{array}
\right]
\end{align}

\subsubsection*{Lifted Hypergraph Products (LHP)}

This column-swapping approach to bias tailoring is broadly applicable. For example, Hadamard rotations can be incorporated directly into the \emph{lifted hypergraph product} (LHP) construction \cite{Roffe2023-ay}. LHP codes follow the same structural blueprint as standard hypergraph product (HGP) codes, but instead of binary parity-check matrices, they take \emph{protographs} as inputs—small symbolic matrices with entries drawn from the \emph{ring of circulants}.

Each ring element \( \lambda_L^\alpha \) represents a rightward cyclic shift of the \( L \times L \) identity matrix by \( \alpha \) positions. For example, \( \lambda_3^1 \) corresponds to the \( 3 \times 3 \) identity matrix with its rows shifted one position to the right:
\vspace{-.25em}
$$\lambda^1_3  = \begin{pmatrix}
    0 & 1 & 0 \\
    0 & 0 & 1 \\
    1 & 0 & 0
\end{pmatrix}$$. 
\vspace{-.25em}

Each ring entry expands into an \( L \times L \) binary matrix, where the 
lift parameter \(L\) specifies the block size. The use of circulants preserves structure, sparsity, and decoding efficiency while allowing algebraic design flexibility at the protograph level.

A matrix composed of these ring elements produces a protograph. For example, the protograph:
\[
A_L =
\begin{bmatrix}
\lambda^1_L + \lambda^2_L & \lambda^0_L & 0 \\
0 & \lambda^0_L + \lambda^1_L & \lambda^1_L
\end{bmatrix}
\]
defines a compact, high-level representation of a parity-check matrix \cite{Roffe2023-ay}. When expanded with lift parameter \(L = 3\), each entry becomes a sum of permutation matrices, and the full protograph expands into a binary matrix of size \(6 \times 9\). This lifted binary matrix is shown below:
\[
\mathfrak{B}(A_3) =
\left[
\begin{array}{ccc|ccc|ccc}
0 & 1 & 1 & 1 & 0 & 0 & 0 & 0 & 0 \\
1 & 0 & 1 & 0 & 1 & 0 & 0 & 0 & 0 \\
1 & 1 & 0 & 0 & 0 & 1 & 0 & 0 & 0 \\
\hline
0 & 0 & 0 & 1 & 1 & 0 & 0 & 1 & 0 \\
0 & 0 & 0 & 0 & 1 & 1 & 1 & 0 & 1 \\
0 & 0 & 0 & 1 & 0 & 1 & 1 & 0 & 0
\end{array}
\right].
\]

\renewcommand{\arraystretch}{2} 
\setlength{\tabcolsep}{8pt}       

\begin{table*}
\caption{Comparison of Hypergraph Product and Lifted Product quantum LDPC constructions.
Here \(n_i\) and \(r_i\) denote the column and row counts of seeds, and \(L\) is the circulant size.}
\label{tab:hgp_vs_lp}

\begin{tabular}{|l|l|l|}
\hline
\textbf{Feature} & \textbf{Hypergraph Product (HGP)} & \textbf{Lifted Product (LP)} \\
\hline
Construction idea
& \makecell[l]{Tensor product of binary parity-check \\ matrices $H_1$, $H_2$}
& \makecell[l]{Tensor product of quasi-cyclic \\ protographs $A_1$, $A_2$} \\
\hline
Blocklength $N$ (lift $L$)
& \makecell[l]{$N_{\text{HGP}} \approx (n_1 n_2 + r_1 r_2)\,L^2$ (quadratic in $L$)}
& \makecell[l]{$N_{\text{LP}} \approx (n_1 + r_1)(n_2 + r_2)\,L$ (linear in $L$)} \\
\hline
Rate $K/N$
& \makecell[l]{$O(1/L)$ for symmetric seeds \\ (vanishing rate)}
& $O(1)$ rate possible with asymptotically good seeds \\
\hline
Distance $d$
& Typically $O(\sqrt{N})$
& $O(N/\mathrm{polylog}\,N)$ \\
\hline
Short-cycle control
& \makecell[l]{Limited (4-cycles from tensor structure)}
& \makecell[l]{Tunable via circulant shifts (girth/ACE)} \\
\hline
Hardware friendliness
& No special structure
& \makecell[l]{Quasi-cyclic layout; regular routing} \\
\hline
\end{tabular}

\end{table*}

\renewcommand{\arraystretch}{1.0}
\setlength{\tabcolsep}{6pt}

This transformation from symbolic protographs to binary matrices retains structure and sparsity while enabling systematic design of large, high-performance LDPC codes.  In particular, the LHP construction we employ generalizes the classic HGP by taking the tensor product of quasi-cyclic protographs—matrices whose entries are circulant permutation blocks—rather than plain binary parity-check matrices (see Table~\ref{tab:hgp_vs_lp}). 

\vspace{0.3em}\noindent
\textbf{Scalability.}  
Because each circulant has size~\(L\), the LHP blocklength grows only linearly with~\(L\), whereas the HGP blocklength scales \(\sim L^{2}\) \cite{Panteleev2022-qh,Tillich2014-cj}.  For large lifts this halves the qubit overhead relative to an HGP code with comparable distance.

\textbf{Rate and distance.}  
LHP codes can maintain a constant rate when the seed protographs \(\{A_{1},A_{2}\}\) are chosen from an asymptotically good family, and their minimum distance scales almost linearly, \(d=\Theta\!\bigl(N/\mathrm{polylog}\,N\bigr)\) \cite{Panteleev2022-bp}.  By contrast, HGP codes of equal-sized seeds have rate \(O(1/L)\) and distance \(\Theta(\sqrt{N})\) \cite{Tillich2014-cj}.

\textbf{Cycle and trapping-set suppression.}  
The circulant shift parameters can be tuned to maximize girth or satisfy ACE constraints, thereby suppressing 4-cycles and other harmful topologies \cite{Bazarsky2013ACE}.  This “short-cycle control’’ gives LHP codes a markedly lower error floor than naive HGP lifts.

\textbf{Hardware friendliness.}  
The quasi-cyclic layout naturally tiles on 2D lattices and admits regular interconnect routing, easing physical mapping and decoder scheduling \cite{Bravyi2024-am,Roffe2023-ay}.

These combined advantages—linear-in-\(L\) scaling, constant rate, tunable short-cycle spectrum, and hardware-oriented structure—make LP-based lifted homological-product codes an attractive platform for bias-tailored single-shot decoding.

Given the seed protographs $A_1$ and $A_2$, the lifted parity-check matrices are defined as:
\begin{align}
A &= [A_X \;|\; A_Z] \notag \\
&= \left[
\begin{array}{cc|cc}
0 & 0 & \mathbb{I} \otimes A_2 & A_1^T \otimes \mathbb{I} \\
A_1 \otimes \mathbb{I} & \mathbb{I} \otimes A_2^T & 0 & 0
\end{array}
\right]
\end{align}

These matrices are then lifted to binary form using a chosen lift parameter $L$, where each protograph entry expands into an $L \times L$ circulant matrix. The $H_X$ and $H_Z$ matrices are derived by lifting $A_X$ and $A_Z$ into their binary forms. Like the hypergraph product for binary matrices, the lifted product enables quantum code construction from arbitrary pairs of protographs \cite{Roffe2023-ay}. 

Critically, bias tailoring can be applied directly at the protograph level: a Hadamard rotation on half the qubits corresponds to swapping symbolic blocks within the protograph. After this transformation, the lifted product matrix takes the following form:
\vspace{-.5em}
\begin{align}
A' &= [A_X \;|\; A_Z] \notag \\
&= \left[
\begin{array}{cc|cc}
0 & A_1^T \otimes \mathbb{I} & \mathbb{I} \otimes A_2 & 0 \\
A_1 \otimes \mathbb{I} & 0 & 0 & \mathbb{I} \otimes A_2^T
\end{array}
\right]
\end{align}

Once again, this transformation yields mixed-type stabilizers that are naturally adapted to biased noise channels, particularly those dominated by dephasing errors. However, unlike conventional HGP codes built from arbitrary binary parity-check matrices, this construction uses structured protographs as seeds. By working at the symbolic level—before expansion into large binary matrices—the lifted hypergraph product (LHP) framework preserves algebraic structure that facilitates more efficient and scalable code design.

Notably, LHP codes derived from protographs often exhibit significantly better parameters than their binary HGP counterparts. They tend to achieve a higher rate (\(k/n\)), larger minimum distance, and improved decoding performance under belief propagation with ordered statistics decoding (BP+OSD). This advantage arises from the quasi-cyclic structure and controlled sparsity of protograph-based codes, which promote better girth, regularity, and convergence properties during iterative decoding \cite{Roffe2023-ay}.
\vspace{-1em}
\section{Methodology}
\label{sec:methods}

\begin{figure}
  \includegraphics[width=\linewidth]{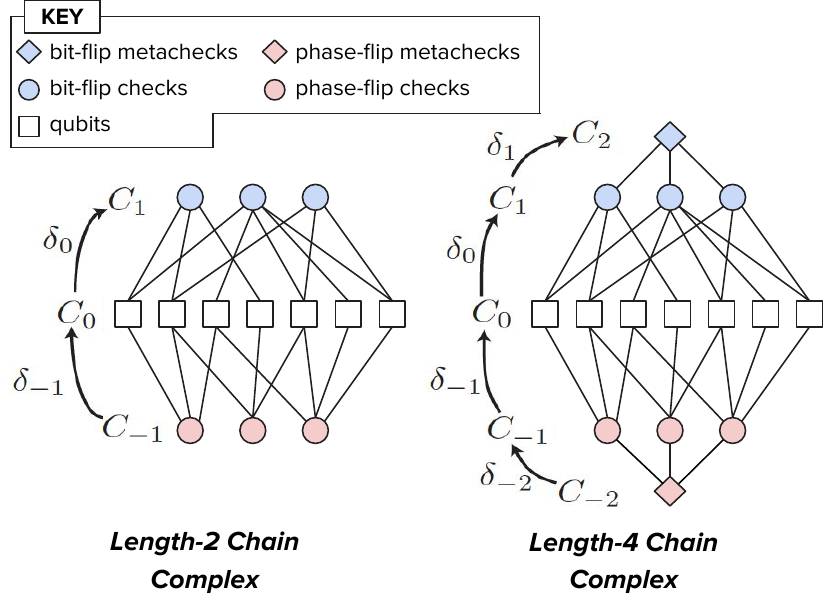}
  \caption{
  \textbf{Left}: a standard CSS code corresponds to a length-2 complex, where qubit errors are detected by bit- and phase-flip checks ($\delta_{\pm1}$). \textbf{Right}: a length-4 complex introduces metachecks ($\delta_{\pm2}$) that validate the syndrome bits themselves. Adapted from \cite{Campbell2019-ez}.} \vspace{-1.5em}
  \label{fig:ss_viz}
\end{figure}

\begin{figure*}[!htbp]
\centering
  \includegraphics[width=\textwidth]{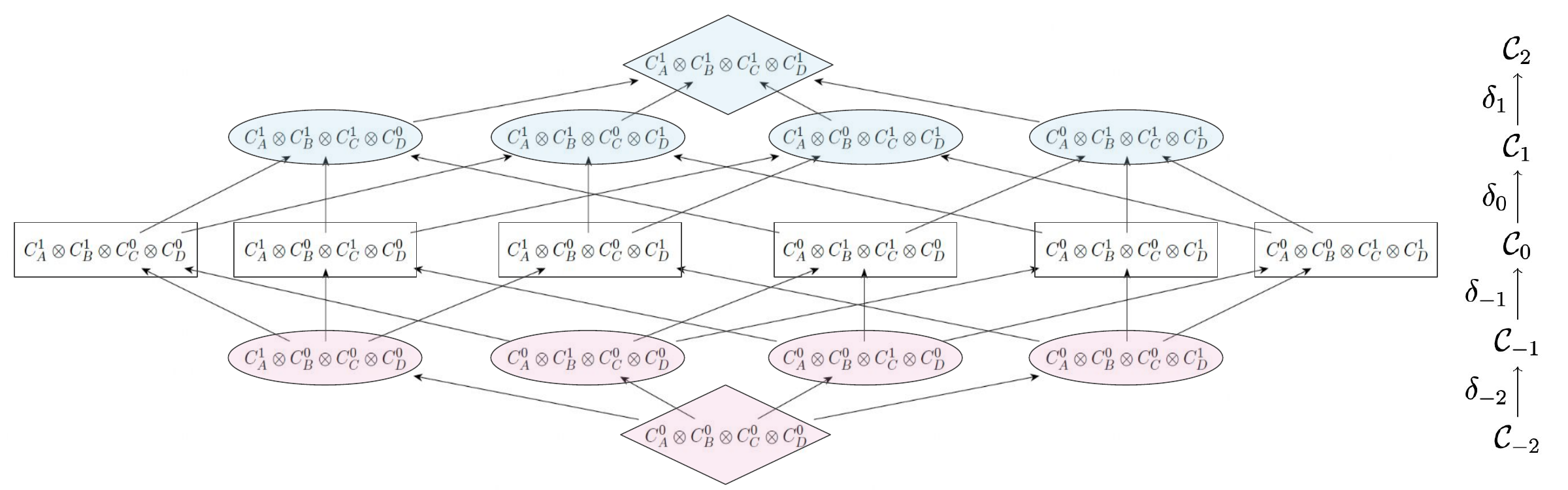}
  \caption{4D lifted chain complex. Each node is a tensor product \( C_A^i \otimes C_B^j \otimes C_C^k \otimes C_D^l \), where \( C^0 \) and \( C^1 \) denote code and check spaces. Nodes are grouped by total degree into five homological layers, with diamonds (\( \mathcal{C}_{\pm2} \)) representing metachecks, ovals (\( \mathcal{C}_{\pm1} \)) representing stabilizers, and rectangles (\( \mathcal{C}_0 \)) representing qubits. Arrows denote boundary maps \( \delta_i \) linking adjacent layers. Blue nodes correspond to $X$-type (bit-flip) operators, and pink nodes to $Z$-type (phase-flip) operators. The structure enables detection and correction of both data and measurement errors via single-shot decoding.
} 
  \label{fig:4d_chain}
\end{figure*}

\subsection{4D Chain Complex}\vspace{-.5em}
This work expands the two-dimensional HGP construction to four dimensions by introducing \emph{four} classical seed matrices, \(\delta_A,\delta_B,\delta_C,\delta_D\). See Fig.~\ref{fig:ss_viz} for a visualization of the multi-dimensional chain complexes and the relationships induced by the boundary maps between adjacent layers. 

Tensor-embedding these seeds produces the length-four chain complex. These matrices generate a length-four chain complex (details shown in Fig.~\ref{fig:4d_chain}). Each node in the diagram represents a tensor product of classical code components of the form
\[
C_A^i \otimes C_B^j \otimes C_C^k \otimes C_D^l,
\]
where each \( C_X^0 \) denotes the code space (primal) and \( C_X^1 \) the parity-check space (dual) of a classical seed code \( C_X \), with \( X \in \{A, B, C, D\} \). The superscripts \( i, j, k, l \in \{0,1\} \) indicate whether each factor is a codeword space or a check space. The total degree \( i + j + k + l \) determines the vertical position of the node in the complex.

Edges between nodes correspond to boundary maps—linear transformations defined via Kronecker products of identity matrices and seed code maps \( \delta_X \). These maps link adjacent layers and encode the syndrome relationships used for error detection.

Each space $\mathcal{C}_i$ corresponds to a distinct role in the code: $\mathcal{C}_0$ encodes the physical qubits, $\mathcal{C}_{\pm1}$ represent the $X$- and $Z$-type stabilizers, and $\mathcal{C}_{\pm2}$ define metachecks that validate the syndromes:
\begin{equation}
\mathcal{C}_{-2} \xrightarrow{\delta_{-2}} \mathcal{C}_{-1} \xrightarrow{\delta_{-1}} \mathcal{C}_0 \xrightarrow{\delta_0} \mathcal{C}_1 \xrightarrow{\delta_1} \mathcal{C}_2
\end{equation}

This layered structure introduces redundancy at the data and measurement levels. Each physical error activates multiple stabilizers, and each measurement error alters the syndrome in a way that triggers multiple metachecks. Thus, the 4D complex supports single-shot decoding: both data and measurement errors can be corrected after a single round of noisy syndrome extraction. \vspace{-2em}
\subsection{4D Matrix Construction}
\vspace{-1em}
To prepare for constructing the full 4D chain complex, first define the identity-expanded forms of the classical seed matrices:

\begin{align}
\dA &:= \delta_A \otimes \id \otimes \id \otimes \id \\
\dB &:= \id \otimes \delta_B \otimes \id \otimes \id \\
\dC &:= \id \otimes \id \otimes \delta_C \otimes \id \\
\dD &:= \id \otimes \id \otimes \id \otimes \delta_D
\end{align} 

Each \(\delta_X\) is the parity-check matrix of a classical seed code \(C_X\), and its expansion via identity tensors ensures that every term acts on the full tensor product space \(C_A \otimes C_B \otimes C_C \otimes C_D\). This uniform embedding is essential for composing boundary maps, where dimensional consistency across terms is required. These identity-augmented forms ensure that all boundary operators are well-defined and structurally aligned within the chain complex.

The resulting boundary maps naturally define the parity-check and metacheck matrices used for syndrome extraction and validation:

\begin{equation}
\delta_{-2}^T = M_Z = 
\begin{pmatrix}
\dA & 
\dB & 
\dC & 
\dD
\end{pmatrix}
\end{equation}

\begin{align}
\delta_{-1}^T = H_Z &= [H_{Z_1} \mid H_{Z_2}] \notag \\
&= 
\begin{pNiceArray}{ccc:ccc}[columns-width=auto,cell-space-limits=2pt]
\dB & \dC & \dD & 0   & 0   & 0 \\
\dA & 0   & 0   & \dC & \dD & 0 \\
0   & \dA & 0   & \dB & 0   & \dD \\
0   & 0   & \dA & 0   & \dB & \dC
\end{pNiceArray}
\end{align}

\begin{align}
\delta_0 = H_X &= [H_{X_1} \mid H_{X_2}] \notag \\
&= 
\begin{pNiceArray}{ccc:ccc}[columns-width=auto,cell-space-limits=2pt]
\dC & \dB & 0   & \dA & 0   & 0 \\
\dD & 0   & \dB & 0   & \dA & 0 \\
0   & \dD & \dC & 0   & 0   & \dA \\
0   & 0   & 0   & \dD & \dC & \dB
\end{pNiceArray}
\end{align}

\begin{equation}
\begin{array}{c}
\delta_1 = M_X = 
\begin{pmatrix}
\dD & \dC & \dB & \dA
\end{pmatrix}
\end{array}
\end{equation}

This construction ensures algebraic consistency ($\delta_i \circ \delta_{i-1} = 0$) and enables the detection of both qubit-level and measurement-level faults. Let the full parity-check matrix be organized as:
\begin{align}
H &= [H_X \mid H_Z] = [H_{X_1} \dashedmid H_{X_2} \mid H_{Z_1} \dashedmid H_{Z_2}]  \\
  &= 
\begin{pNiceArray}{ccc:ccc|ccc:ccc}[columns-width=auto,cell-space-limits=2pt]
\dC & \dB & 0   & \dA & 0   & 0   & \dB & \dC & \dD & 0   & 0   & 0 \\
\dD & 0   & \dB & 0   & \dA & 0   & \dA & 0   & 0   & \dC & \dD & 0 \\
0   & \dD & \dC & 0   & 0   & \dA & 0   & \dA & 0   & \dB & 0   & \dD \\
0   & 0   & 0   & \dD & \dC & \dB & 0   & 0   & \dA & 0   & \dB & \dC
\end{pNiceArray} \notag 
\end{align}

As with the 2D HGP construction (Eq.~\ref{hgp}, Eq.~\ref{hgp_swap}), the Hadamard rotation is implemented by swapping the rightmost halves of \( H_X \) and \( H_Z \). 
\begin{align}
H' &= [H_{X_1} \dashedmid H_{Z_2} \mid H_{Z_1} \dashedmid H_{X_2}]  \\
   &= 
\begin{pNiceArray}{ccc:ccc|ccc:ccc}[columns-width=auto,cell-space-limits=2pt]
\dC & \dB & 0   & 0   & 0   & 0   & \dB & \dC & \dD & \dA & 0   & 0 \\
\dD & 0   & \dB & \dC & \dD & 0   & \dA & 0   & 0   & 0   & \dA & 0 \\
0   & \dD & \dC & \dB & 0   & \dD & 0   & \dA & 0   & 0   & 0   & \dA \\
0   & 0   & 0   & 0   & \dB & \dC & 0   & 0   & \dA & \dD & \dC & \dB
\end{pNiceArray}  \notag 
\end{align}

After this transformation, both \( H_X \) and \( H_Z \) exhibit centrosymmetric structure: the right half is a 180-degree rotation of the left. This symmetry aligns the code with the dephasing-biased noise model while preserving commutation between stabilizers \cite{Akarsu2022-vw}.

\vspace{-.5em}
\subsection{Biased LHP Code Simulation}
\vspace{-.5em}
To implement the 4D-LHP code construction, we extended functionality from two core libraries: the \texttt{LDPC} package, which provides a  protograph framework, and the \texttt{BPOSD} package, which defines CSS and general stabilizer code classes \cite{ldpc_repo, bposd_repo}. 

To incorporate lifting, our \texttt{LHP4D} class takes symbolic protographs as input matrices $(\delta_A, \delta_B, \delta_C, \delta_D)$ and performs intermediate operations directly in the symbolic domain before compiling the final parity-check matrices to binary form via the lift parameter $L$. We also adapted design patterns and techniques from the \texttt{Bias Tailored qLDPC} repository \cite{bias_tailored_repo}, which served as a practical reference for implementing LHP codes.

Bias tailoring is applied dynamically during simulation by modifying the error model rather than altering the code structure itself. The qubits are partitioned into two sectors corresponding to halves of the $H_X$ and $H_Z$ matrices. The first sector is not rotated, so it retains the standard Pauli error probabilities. In the second sector, the roles of $X$ and $Z$ are interchanged, meaning qubits that would typically receive $Z$ errors are now subject to $X$ errors (and vice versa). 

This swap simulates the action of a Hadamard transformation on those qubits. The resulting channel asymmetry enables the simulation of bias-aware decoding without altering the code structure itself, preserving compatibility with standard CSS decoders.
\vspace{-.5em}
\subsection{Noise Model and Single-Shot Decoding}

The SS decoding strategy used in this work consists of two decoding stages and a failure test, described in Algorithm~\ref{alg:decode}. This approach draws on the protocol implemented in the \texttt{single\_shot\_3D\_HGP} repository \cite{Vasmer2023-single-shot}.

To evaluate the performance of bias-tailored LHP codes, we performed numerical simulations under a circuit-level noise model. Each physical qubit experiences Pauli noise with total probability $p$. Noise bias is introduced via parameters $\beta_X, \beta_Y, \beta_Z$, which define the relative rates of each Pauli error. The corresponding error probabilities are computed as:
\begin{equation}
p_X = \frac{p \cdot \beta_X}{\beta}, \quad
p_Y = \frac{p \cdot \beta_Y}{\beta}, \quad
p_Z = \frac{p \cdot \beta_Z}{\beta}, \quad 
\end{equation}
\vspace{-1.4em}
$$
\text{where } \beta = \beta_X + \beta_Y + \beta_Z.
$$

In addition to data qubit errors, each stabilizer measurement may independently fail with probability $q$. A measurement error results in a flipped syndrome bit, potentially masking or mimicking a physical error. To decode under this joint data-and-syndrome noise model, we employ the two-step single-shot decoding strategy \cite{Quintavalle2021-hr}. 

In the first stage, the observed syndrome—potentially corrupted by measurement noise—is interpreted as a codeword in the metacheck code and decoded using BP+OSD to locate and correct any measurement errors. The resulting corrected syndrome is then passed to a second BP+OSD decoder, which infers the corresponding physical error. If both decoding steps succeed, the combined recovery operator restores the system to the codespace.

After decoding, the residual error is computed as the sum of the true error and the inferred recovery operator:
\begin{equation}
\varepsilon_X = e_X + \hat{e}_X, \quad \varepsilon_Z = e_Z + \hat{e}_Z
\end{equation}
The residual represents the net remaining error after correction. To determine whether the decoding has succeeded, the residual error is tested against all logical operators. Specifically, a logical failure is declared if the residual error anticommutes with any logical operator:
\begin{equation}
(L_Z \cdot \varepsilon_X) \vee (L_X \cdot \varepsilon_Z) \neq 0,
\end{equation}
\noindent as this failure indicates that the error has induced a nontrivial logical operation on the encoded state. Otherwise, decoding is deemed successful, ensuring that the recovery operator corrects the syndrome and preserves the encoded logical information. 
\vspace{-.5em}
\subsection{Protograph Seed Selection and Code Parameters}\label{sec:peg}
\vspace{-.5em}

From several candidates, we selected a small-scale protograph with high local girth and no weight-two cycles:
\begin{equation}
\begin{bmatrix}
\lambda(2) & \lambda() & \lambda(0) & \lambda(0) & \lambda() & \lambda(2) \\
\lambda()  & \lambda(0) & \lambda(2) & \lambda() & \lambda(2) & \lambda(0) \\
\lambda(0) & \lambda(1) & \lambda() & \lambda(1) & \lambda(0) & \lambda(2) \\
\lambda(0) & \lambda(1) & \lambda(1) & \lambda(1) & \lambda(1) & \lambda()
\end{bmatrix}
\end{equation}

Lifting with $L = 3$ yields a quantum code with parameters $[[384, 48, 6]]$ (rate $K/N = 0.125$).  This code is competitive with state-of-the-art LDPC designs. Table~\ref{tab:code_comparison} compares this representative 4D-LHP instance with widely studied baselines at similar distance. For $L{=}3$, the construction achieves $[[384,48,6]]$ (rate $0.125$) while maintaining bounded stabilizer weight and metacheck redundancy.


\begin{table}[h]
\centering
\caption{Comparison of codes at distance $d=6$}
\label{tab:code_comparison}
\begin{tabular}{lccc}
\hline
\textbf{Code} & \textbf{Rate $(k/n)$} & \textbf{Qubits $(n)$} & \textbf{Avg. Weight} \\
\hline
4D-LHP (Ours) & 0.125 & 384 & 5.2 \\
HGP~\cite{Tillich2014-cj} & 0.04 & 400 & 6.1 \\
Surface Code & 0.11 & 1134 & 4.0 \\
\hline
\end{tabular}
\end{table}

Against this backdrop, the finite-length rate of 0.125 demonstrates that the 4D–LHP architecture can match or exceed the efficiency of much larger asymptotic codes, while retaining practical properties such as bounded stabilizer weight and support for single-shot decoding.

\vspace{-.75em}
\begin{figure}
  \includegraphics[width=\linewidth]{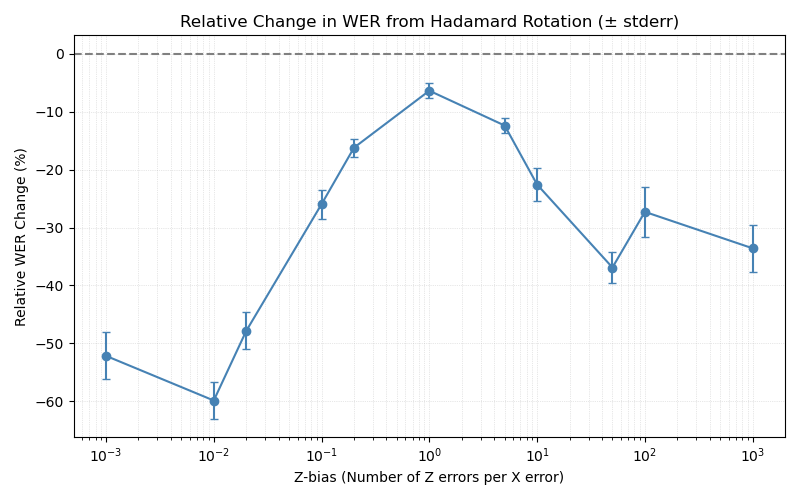}
  \caption{ Benefit of Hadamard bias-tailoring. Relative change in WER (negative values indicate improvement) as a function of the $Z{:}X$ bias ratio $\eta=p_Z/p_X$. Error bars show ± 1 s.e.m. Bias tailoring yields greatest gains (up to 60 \%) in moderate bias $10^{-3}\!\le\!\eta\!\le\!10^{-1}$.  
}
  \label{fig:hrot}
\end{figure}

\begin{figure*}
  \includegraphics[width=\linewidth]{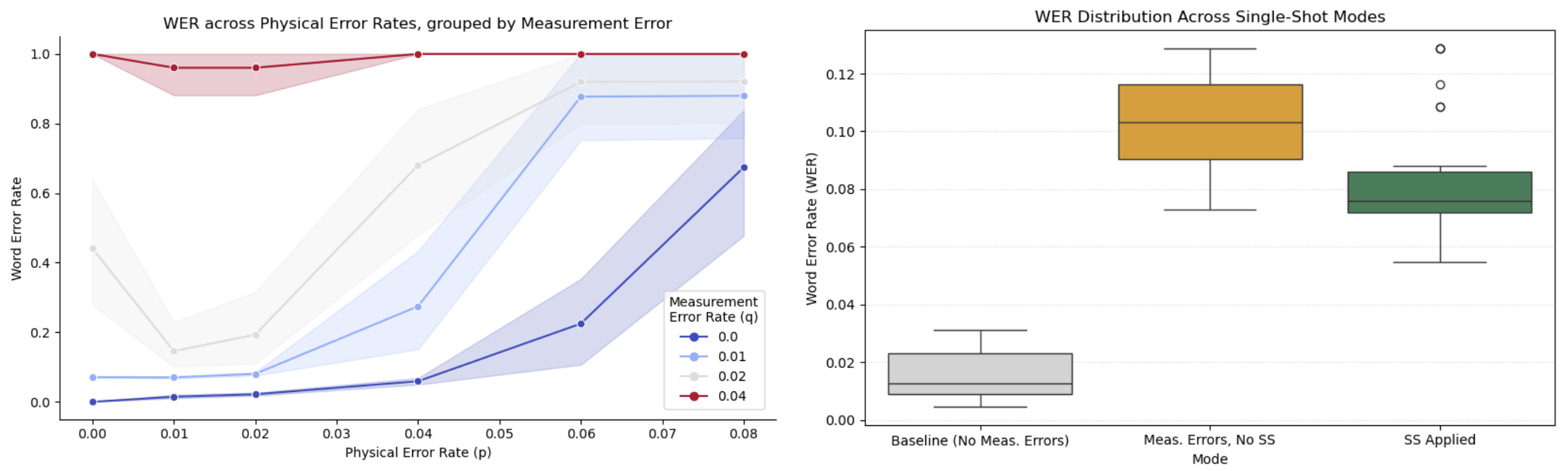}
  \caption{Effect of measurement errors and single-shot (SS) decoding on word-error rate (WER).  
  \textbf{Left}:  WER vs.\ physical error rate \(p\) for several measurement-error rates \(q\) (no SS decoding). Even modest \(q\ge 0.02\) causes an order-of-magnitude rise in logical failures.  
    \textbf{Right}: WER distributions at a representative error rate (\(p=0.04\)). 
    Shown are the baseline (perfect measurements), noisy measurements without SS, 
    and noisy measurements with SS. A single round of single-shot decoding lowers 
    the median WER from 10.3\% to 7.4\% (28\% reduction). Relative to the 1.6\% 
    baseline, this recovers over one-third of the penalty introduced by measurement 
    noise.
    } \vspace{-1em}
  \label{fig:ss_combo}
\end{figure*}

\section{Results}
\label{sec:results}

To quantify performance, the simulation measures the \emph{word error rate} (WER), defined as the probability that the decoder fails to return the correct logical state. Monte Carlo trials were conducted over a range of physical error rates \(p \in [0.01, 0.08]\), measurement error rates \(q \in [0.0, 0.04]\), and bias ratios \(\eta = p_Z / p_X\) ranging from \(10^{-3}\) to \(10^3\).

Figure~\ref{fig:hrot} reports the \emph{relative} change in WER induced by bias tailoring, expressed as a percentage difference between the untailored and Hadamard-rotated code. Across the entire range of tested error rates and noise biases, the Hadamard rotation  improved performance—lowering the WER in every tested configuration.

The largest gains were observed in the moderately biased regime (\(10^{-2} \le \eta \le 10^{-1}\)), where WER dropped by up to 60\%. Even in the near-unbiased case (\(\eta \approx 1\)), the tailored code outperformed its untwisted counterpart by approximately 10\%, indicating that the rotation does not degrade performance even when bias is weak. For extremely high bias (\(\eta \gg 10^2\)), the benefit remains substantial (up to 40\%), though it tapers slightly—likely due to error patterns concentrating in highly confined sectors where decoder performance saturates.

These results confirm that bias tailoring via a Hadamard rotation consistently enhances the code’s resilience to physical noise by increasing the effective distance of dominant-error logical operators. Notably, this improvement holds regardless of the underlying error rates \(p\) and  \(q\), making the technique broadly applicable across realistic NISQ regimes.

Figure~\ref{fig:ss_combo}a plots the WER of the 4D-LHP code as a function of the physical error rate \(p\), stratified by measurement error probability \(q\). In the ideal-measurement limit (\(q = 0\)), the decoder maintains high reliability: the WER remains below \(10^{-2}\) for physical error rates up to \(p \simeq 0.06\), and remains below 0.1 even at \(p = 0.08\). This performance baseline reflects the effectiveness of the BP+OSD decoder under purely data-driven noise.

However, once measurement noise is introduced, performance degrades rapidly. For \(q = 0.01\), the WER begins to rise sharply beyond \(p = 0.04\), and for \(q = 0.02\), logical errors become frequent across the full range of tested \(p\) values. At \(q = 0.04\), decoding fails almost deterministically, with WER saturating near 1.0 for all values of \(p\). These results highlight the vulnerability of quantum LDPC codes to even modest levels of measurement noise if left uncorrected.

The increasing spread between the curves as \(p\) increases underscores the compounding effect of data and measurement errors. In the absence of metachecks, syndrome faults propagate directly into decoding failures, exacerbating the logical error rate. These findings underscore the necessity of explicit mechanisms for correcting measurement errors—such as single-shot decoding—when operating under realistic, noisy readout conditions.

Notably, the lowest measurement noise configuration (\(q = 0.0\)) defines an upper bound on achievable performance for the given code and decoder, serving as a target baseline for evaluating the effectiveness of single-shot recovery in the next section.

Figure~\ref{fig:ss_combo}b compares the distribution of word error rates (WER) across three decoding modes:  
(i) a baseline configuration with noiseless measurement ,  
(ii) decoding under realistic measurement noise without single-shot (SS) correction, and  
(iii) decoding under measurement noise with the SS layer applied.

The baseline mode establishes the optimal performance expected in the absence of measurement faults, with a median WER below \(0.02\) and tight variance across trials. Introducing measurement errors without correction degrades performance sharply: the median WER rises to \(11 \% \), and the interquartile range expands significantly, indicating unstable and unreliable decoding. This degradation underscores the critical role that accurate syndrome extraction plays in quantum LDPC decoding.

When the single-shot layer is enabled, the decoder regains a substantial portion of lost performance. The median WER drops to \(\sim0.08\), recovering roughly one third of the performance gap caused by measurement noise. 

Together, these results validate the efficacy of single-shot decoding in practical, noisy scenarios. Although the SS pipeline does not fully restore ideal performance, the metacheck layer meaningfully reduces the logical failure rate and narrows performance variation, offering a practical pathway toward fault-tolerant operation.

\begin{figure}
  \includegraphics[width=\linewidth]{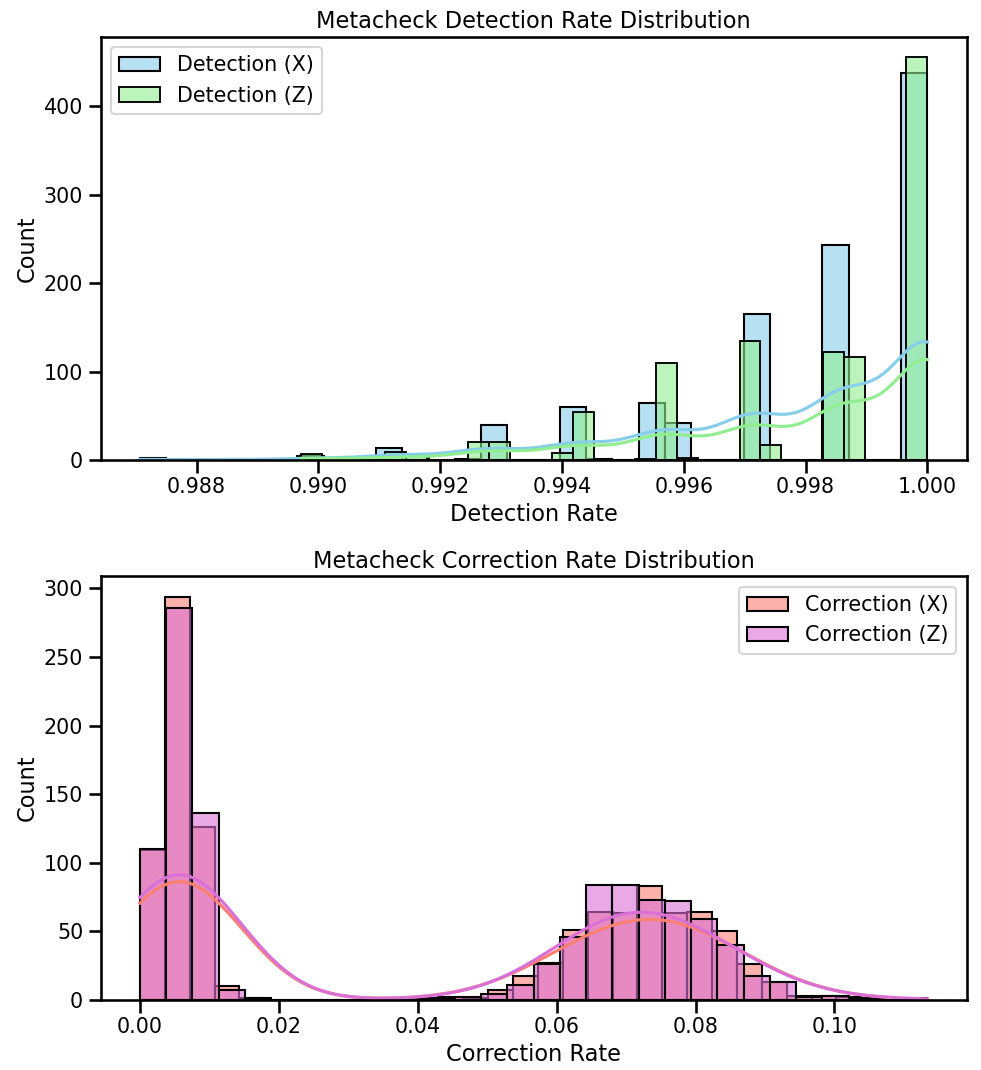}
  \caption{\textbf{Top panel: Detection rates} for $X$-type (blue) and $Z$-type (green) measurement faults concentrate tightly near 1.0, confirming near-perfect fault identification. \textbf{Bottom panel: Correction rates} are more dispersed, peaking around 7-8\%, with a notable spike near zero for both error types. Although far fewer faults are corrected than detected, this limited success is sufficient to significantly reduce WER. 
} \vspace{1em}
  \label{fig:metacheck}
\end{figure}

Evaluating metacheck performance is essential for assessing the single-shot capabilities of the current code and for guiding the design of optimized 4D-LHP codes. Figure~\ref{fig:metacheck} analyzes the effectiveness of the metacheck layer by reporting the distribution of detection and correction rates across Monte Carlo trial.

Detection rates are tightly concentrated near \(1.0\) for both $X$- and $Z$-type measurement faults. Most trials achieve fault identification accuracy above \(99.8\%\), confirming that the metasyndrome reliably signals the presence of measurement errors. The sharp rightward skew in the histogram indicates that near-perfect detection is not an outlier but rather the dominant regime. This highlights the structural redundancy of the 4D chain complex: each measurement error affects multiple overlapping metachecks, boosting the likelihood of detection.

However, detection does not guarantee correction. The metacheck correction rate—the proportion of measurement faults successfully corrected—is more broadly distributed. While many trials achieve correction rates in the 5–10\% range, a substantial fraction see near-zero correction. This bimodal distribution indicates that correction success depends heavily on the specific fault pattern and decoder convergence. Nonetheless, even partial correction has meaningful impact: as shown in Fig.~\ref{fig:ss_combo}b, correcting just a small subset of faulty syndromes is sufficient to materially reduce the word error rate.

Together, these results validate the design of the metacheck layer. It provides nearly full fault \textit{visibility} (detection), and enough \textit{actionability} (correction) to recover significant performance. Improving the correction success rate—e.g., through tailored seed selection or metacheck-specific decoding heuristics—could further amplify the benefit of single-shot decoding in high-noise or bias-skewed regimes.

Across the simulated parameter space, the 4D-LHP code exhibits:
\begin{enumerate}
  \item \textbf{Strong bias responsiveness:} a tailored Hadamard rotation cuts logical errors by up to \(60\%\) in regimes of moderate asymmetry.
  \item \textbf{Robustness to measurement noise:} single-shot decoding restores a substantial portion of baseline performance at realistic measurement-error rates.
  \item \textbf{High fault visibility:} metachecks detect almost all measurement faults, and their partial correction materially improves overall reliability.
\end{enumerate}

These results demonstrate that combining bias tailoring with single-shot decoding in a 4D lifted-product architecture yields significant practical advantages under circuit-level noise.
\vspace{-1em}
\section{Discussion}
\label{sec:disc}
\vspace{-1em}
The numerical results presented above illustrate how a single architectural framework---a
four--dimensional lifted homological product (4D~LHP) code equipped with Hadamard
bias-tailoring and single-shot (SS) decoding---addresses \emph{two} practical hurdles that
currently limit near-term quantum devices: biased physical noise and faulty
syndrome extraction.

Below, we first contextualize the empirical gains, then outline concrete avenues
for boosting metacheck effectiveness, and finally chart broader
directions for next-generation code design and decoding.

\subsection{Interpreting the Numerical Gains}
\vspace{-0.5em}

\paragraph*{Additive benefits.}
Bias tailoring and SS decoding have, until now, advanced along mostly
separate lines of inquiry.  The present work shows they can be combined
\emph{constructively}. The tailored Hadamard rotation reduces the word
error rate (WER) by as much as \(60\%\) in the moderately biased regime
(\(\eta \!\sim\! 10^{-2}\); Fig.~\ref{fig:hrot})

In parallel, the SS layer mitigates roughly one third of the logical errors introduced by measurement noise (Fig.~\ref{fig:ss_combo}a). When both techniques are applied, they reduce errors independently: bias tailoring leverages the Hadamard rotation to align with the noise bias, while SS decoding exploits structural redundancy in the chain complex.

\paragraph*{Metacheck visibility vs.\ actionability.}
The 4D complex furnishes near-perfect \emph{visibility} of measurement
faults, as more than \(99.8\%\) of faulty syndromes are
flagged. However, only $5$–$10\%$ of those detected faults are successfully corrected
(Fig.~\ref{fig:metacheck}). This apparent discrepancy is
characteristic of current LDPC decoders, which are highly sensitive to
short cycles and trapping sets \cite{Richardson2001-trapping}.  Even so,
partial correction notably diminishes the WER relative to an architecture
without metachecks. Hence, the bottleneck is not detection but rather \emph{actionability}—a decoder-centric limitation that future work can address.

\subsection{Improving Metacheck Correction Efficiency}
\vspace{-0.5em}

\paragraph*{Seed optimisation with confinement heuristics.}
Detection without correction typically arises when the measurement-error
pattern spans a ``loose’’ metacheck boundary that the BP+OSD decoder
fails to close.  Classical seeds with stronger \emph{confinement}
\cite{Quintavalle2021-hr} yield tighter, low-weight boundaries and
thereby increase the likelihood that noisy syndromes remain correctable under BP+OSD decoding.  Extending progressive-edge growth (PEG) searches to include confinement and trapping-set metrics
\cite{He2018-peg} could therefore increase the metacheck-correction
fraction beyond today’s 10\,\%.

\paragraph*{Decoders that exploit metasyndrome structure.}
Standard BP treats each check on an equal footing; it is oblivious to
the fact that metachecks represent \emph{syndrome} correlations rather
than qubit correlations. Tailored message-passing rules
\cite{Pryadko2020-metabp} or neural decoders trained on the joint space
of data- and meta-syndromes \cite{Varsamopoulos_2017} may extract more
information from the same parity budget, converting visibility into
actionability with modest classical overhead.

\paragraph*{Adaptive two-level scheduling.}
Because metachecks are cheap to measure (they act on classical bits),
one can iterate a light-weight meta-decoding step \emph{in situ}: if a
subset of metachecks remains unsatisfied after the first SS pass, only
those checks need be remeasured, dramatically reducing latency while
probabilistically lifting correction rates toward the detection limit
\cite{Bombin2016-oe}. 

\subsection{Future Directions}
\vspace{-.5em}

\textit{Systematic seed search via PEG.}  
While this work employed PEG-optimized protographs for the seed matrices \(\{\delta_A, \delta_B, \delta_C, \delta_D\}\), the search was heuristic and manually tuned for girth. A more systematic exploration of the seed space—guided by structural metrics tailored to 4D-LHP codes—could uncover configurations with significantly higher metacheck-correction efficiency. 

In particular, extending PEG to explicitly prioritize local confinement (i.e., limiting the spatial spread of error patterns through the Tanner graph) and avoiding known trapping sets may yield seeds that not only support good code distance but also exhibit stronger metasyndrome resolvability. This direction could benefit from combining classical PEG techniques \cite{He2018-peg} with emerging graph-theoretic or machine-learning-guided search heuristics that explore the vast combinatorial space of symbolic protographs.

\textit{Sustained threshold under repeated decoding.}  
This work analyzes code performance in a single-cycle noise model, wherein a round of noise is followed by a single-shot decode. While this is sufficient to validate metacheck functionality, it does not yet establish a true fault-tolerance threshold. In a real fault-tolerant architecture, errors accumulate over many cycles of syndrome extraction, decoding, and correction. 

To assess long-term stability, future simulations must extend to sustained-depth regimes—repeatedly applying circuit-level noise, single-shot decoding, and fault propagation tracking across multiple rounds. Following the methodology of \cite{Quintavalle2021-hr}, one can determine whether logical error rates saturate (threshold behavior) or accumulate unboundedly over time. A favorable sustained threshold would demonstrate that 4D-LHP codes are not only effective in isolated rounds but also suitable for long-duration memory or computation.

\textit{One-stage decoders for metacheck integration.}  
The current decoding approach separates syndrome cleaning and error correction into two stages: first correcting measurement noise using metachecks, then correcting data errors using the cleaned syndrome. While robust, this method incurs classical overhead, potential latency, and dependency between decoders. Recent studies \cite{Higgott2023-hz} suggest that a single-stage belief propagation (BP) decoder—capable of jointly interpreting data and measurement errors—can outperform two-stage decoding, especially in sparse codes with correlated error structures. 

Adapting such a decoder to the 4D-LHP setting would require integrating metachecks directly into the Tanner graph and modifying message-passing rules to reflect the layered structure of the chain complex. If successful, this could significantly reduce decoding latency and memory requirements, making 4D-LHP codes more viable for real-time decoding in hardware-constrained settings.

\section{Conclusion}
\vspace{-.75em}
This work introduces a four-dimensional lifted homological-product (4D-LHP) code that tackles two major challenges in quantum error correction: biased physical noise and faulty stabilizer measurements. By combining a tailored Hadamard rotation with a chain-complex structure that supports metachecks, the code achieves both bias alignment and single-shot (SS) decoding.

Together, these mechanisms reduce the word error rate (WER) by up to an order of magnitude compared to untailored, non-SS baselines—cutting logical failures by 60\% under bias and recovering a third of the performance lost to measurement noise. Importantly, the approach is scalable: the same design principles can be applied to build high-rate, low-overhead LDPC codes suitable for near-term quantum memories.

Looking forward, the integration of bias tailoring and SS decoding into logical gate protocols—such as lattice surgery or twist-based operations—could further reduce overhead in biased-noise processors \cite{Brown2022-bs,Roberts2022-rv}. More broadly, this work illustrates how algebraic code constructions, noise-adapted Clifford transformations, and modern decoding techniques can be combined to meet the practical demands of early fault-tolerant quantum hardware.


\bibliography{bib}

\begin{thebibliography}{33}%
\makeatletter
\providecommand \@ifxundefined [1]{%
 \@ifx{#1\undefined}
}%
\providecommand \@ifnum [1]{%
 \ifnum #1\expandafter \@firstoftwo
 \else \expandafter \@secondoftwo
 \fi
}%
\providecommand \@ifx [1]{%
 \ifx #1\expandafter \@firstoftwo
 \else \expandafter \@secondoftwo
 \fi
}%
\providecommand \natexlab [1]{#1}%
\providecommand \enquote  [1]{``#1''}%
\providecommand \bibnamefont  [1]{#1}%
\providecommand \bibfnamefont [1]{#1}%
\providecommand \citenamefont [1]{#1}%
\providecommand \href@noop [0]{\@secondoftwo}%
\providecommand \href [0]{\begingroup \@sanitize@url \@href}%
\providecommand \@href[1]{\@@startlink{#1}\@@href}%
\providecommand \@@href[1]{\endgroup#1\@@endlink}%
\providecommand \@sanitize@url [0]{\catcode `\\12\catcode `\$12\catcode `\&12\catcode `\#12\catcode `\^12\catcode `\_12\catcode `\%12\relax}%
\providecommand \@@startlink[1]{}%
\providecommand \@@endlink[0]{}%
\providecommand \url  [0]{\begingroup\@sanitize@url \@url }%
\providecommand \@url [1]{\endgroup\@href {#1}{\urlprefix }}%
\providecommand \urlprefix  [0]{URL }%
\providecommand \Eprint [0]{\href }%
\providecommand \doibase [0]{https://doi.org/}%
\providecommand \selectlanguage [0]{\@gobble}%
\providecommand \bibinfo  [0]{\@secondoftwo}%
\providecommand \bibfield  [0]{\@secondoftwo}%
\providecommand \translation [1]{[#1]}%
\providecommand \BibitemOpen [0]{}%
\providecommand \bibitemStop [0]{}%
\providecommand \bibitemNoStop [0]{.\EOS\space}%
\providecommand \EOS [0]{\spacefactor3000\relax}%
\providecommand \BibitemShut  [1]{\csname bibitem#1\endcsname}%
\let\auto@bib@innerbib\@empty
\bibitem [{\citenamefont {Preskill}(2018)}]{Preskill2018-zv}%
  \BibitemOpen
  \bibfield  {author} {\bibinfo {author} {\bibfnamefont {J.}~\bibnamefont {Preskill}},\ }\href@noop {} {\bibfield  {journal} {\bibinfo  {journal} {Quantum}\ }\textbf {\bibinfo {volume} {2}},\ \bibinfo {pages} {79} (\bibinfo {year} {2018})}\BibitemShut {NoStop}%
\bibitem [{\citenamefont {Shor}(1996)}]{Shor1996-ct}%
  \BibitemOpen
  \bibfield  {author} {\bibinfo {author} {\bibfnamefont {P.}~\bibnamefont {Shor}},\ }in\ \href {https://doi.org/10.1109/SFCS.1996.548464} {\emph {\bibinfo {booktitle} {Proceedings of 37th Conference on Foundations of Computer Science}}}\ (\bibinfo {year} {1996})\ pp.\ \bibinfo {pages} {56--65}\BibitemShut {NoStop}%
\bibitem [{\citenamefont {Bomb{\'\i}n}(2015)}]{Bombin2015-sf}%
  \BibitemOpen
  \bibfield  {author} {\bibinfo {author} {\bibfnamefont {H.}~\bibnamefont {Bomb{\'\i}n}},\ }\href@noop {} {\bibfield  {journal} {\bibinfo  {journal} {Phys. Rev. X.}\ }\textbf {\bibinfo {volume} {5}} (\bibinfo {year} {2015})}\BibitemShut {NoStop}%
\bibitem [{\citenamefont {Bombin}(2016)}]{Bombin2016-oe}%
  \BibitemOpen
  \bibfield  {author} {\bibinfo {author} {\bibfnamefont {H.}~\bibnamefont {Bombin}},\ }\href@noop {} {\bibfield  {journal} {\bibinfo  {journal} {Phys. Rev. X.}\ }\textbf {\bibinfo {volume} {6}} (\bibinfo {year} {2016})}\BibitemShut {NoStop}%
\bibitem [{\citenamefont {Quintavalle}\ \emph {et~al.}(2021)\citenamefont {Quintavalle}, \citenamefont {Vasmer}, \citenamefont {Roffe},\ and\ \citenamefont {Campbell}}]{Quintavalle2021-hr}%
  \BibitemOpen
  \bibfield  {author} {\bibinfo {author} {\bibfnamefont {A.~O.}\ \bibnamefont {Quintavalle}}, \bibinfo {author} {\bibfnamefont {M.}~\bibnamefont {Vasmer}}, \bibinfo {author} {\bibfnamefont {J.}~\bibnamefont {Roffe}},\ and\ \bibinfo {author} {\bibfnamefont {E.~T.}\ \bibnamefont {Campbell}},\ }\href@noop {} {\bibfield  {journal} {\bibinfo  {journal} {PRX Quantum}\ }\textbf {\bibinfo {volume} {2}} (\bibinfo {year} {2021})}\BibitemShut {NoStop}%
\bibitem [{\citenamefont {Campbell}(2019)}]{Campbell2019-ez}%
  \BibitemOpen
  \bibfield  {author} {\bibinfo {author} {\bibfnamefont {E.~T.}\ \bibnamefont {Campbell}},\ }\href@noop {} {\bibfield  {journal} {\bibinfo  {journal} {Quantum Sci. Technol.}\ }\textbf {\bibinfo {volume} {4}},\ \bibinfo {pages} {025006} (\bibinfo {year} {2019})}\BibitemShut {NoStop}%
\bibitem [{\citenamefont {Ruiz}\ \emph {et~al.}(2025)\citenamefont {Ruiz}, \citenamefont {Guillaud}, \citenamefont {Leverrier}, \citenamefont {Mirrahimi},\ and\ \citenamefont {Vuillot}}]{Ruiz2025-hx}%
  \BibitemOpen
  \bibfield  {author} {\bibinfo {author} {\bibfnamefont {D.}~\bibnamefont {Ruiz}}, \bibinfo {author} {\bibfnamefont {J.}~\bibnamefont {Guillaud}}, \bibinfo {author} {\bibfnamefont {A.}~\bibnamefont {Leverrier}}, \bibinfo {author} {\bibfnamefont {M.}~\bibnamefont {Mirrahimi}},\ and\ \bibinfo {author} {\bibfnamefont {C.}~\bibnamefont {Vuillot}},\ }\href@noop {} {\bibfield  {journal} {\bibinfo  {journal} {Nat. Commun.}\ }\textbf {\bibinfo {volume} {16}},\ \bibinfo {pages} {1040} (\bibinfo {year} {2025})}\BibitemShut {NoStop}%
\bibitem [{\citenamefont {Puri}\ \emph {et~al.}(2020)\citenamefont {Puri}, \citenamefont {St-Jean}, \citenamefont {Gross}, \citenamefont {Grimm}, \citenamefont {Frattini}, \citenamefont {Iyer}, \citenamefont {Krishna}, \citenamefont {Touzard}, \citenamefont {Jiang}, \citenamefont {Blais}, \citenamefont {Flammia},\ and\ \citenamefont {Girvin}}]{Puri2020-vm}%
  \BibitemOpen
  \bibfield  {author} {\bibinfo {author} {\bibfnamefont {S.}~\bibnamefont {Puri}}, \bibinfo {author} {\bibfnamefont {L.}~\bibnamefont {St-Jean}}, \bibinfo {author} {\bibfnamefont {J.~A.}\ \bibnamefont {Gross}}, \bibinfo {author} {\bibfnamefont {A.}~\bibnamefont {Grimm}}, \bibinfo {author} {\bibfnamefont {N.~E.}\ \bibnamefont {Frattini}}, \bibinfo {author} {\bibfnamefont {P.~S.}\ \bibnamefont {Iyer}}, \bibinfo {author} {\bibfnamefont {A.}~\bibnamefont {Krishna}}, \bibinfo {author} {\bibfnamefont {S.}~\bibnamefont {Touzard}}, \bibinfo {author} {\bibfnamefont {L.}~\bibnamefont {Jiang}}, \bibinfo {author} {\bibfnamefont {A.}~\bibnamefont {Blais}}, \bibinfo {author} {\bibfnamefont {S.~T.}\ \bibnamefont {Flammia}},\ and\ \bibinfo {author} {\bibfnamefont {S.~M.}\ \bibnamefont {Girvin}},\ }\href@noop {} {\bibfield  {journal} {\bibinfo  {journal} {Sci. Adv.}\ }\textbf {\bibinfo {volume} {6}},\ \bibinfo {pages} {eaay5901} (\bibinfo {year} {2020})}\BibitemShut {NoStop}%
\bibitem [{\citenamefont {Bonilla~Ataides}\ \emph {et~al.}(2021)\citenamefont {Bonilla~Ataides}, \citenamefont {Tuckett}, \citenamefont {Bartlett}, \citenamefont {Flammia},\ and\ \citenamefont {Brown}}]{Bonilla_Ataides2021-id}%
  \BibitemOpen
  \bibfield  {author} {\bibinfo {author} {\bibfnamefont {J.~P.}\ \bibnamefont {Bonilla~Ataides}}, \bibinfo {author} {\bibfnamefont {D.~K.}\ \bibnamefont {Tuckett}}, \bibinfo {author} {\bibfnamefont {S.~D.}\ \bibnamefont {Bartlett}}, \bibinfo {author} {\bibfnamefont {S.~T.}\ \bibnamefont {Flammia}},\ and\ \bibinfo {author} {\bibfnamefont {B.~J.}\ \bibnamefont {Brown}},\ }\href@noop {} {\bibfield  {journal} {\bibinfo  {journal} {Nat. Commun.}\ }\textbf {\bibinfo {volume} {12}},\ \bibinfo {pages} {2172} (\bibinfo {year} {2021})}\BibitemShut {NoStop}%
\bibitem [{\citenamefont {Tillich}\ and\ \citenamefont {Zemor}(2014)}]{Tillich2014-cj}%
  \BibitemOpen
  \bibfield  {author} {\bibinfo {author} {\bibfnamefont {J.-P.}\ \bibnamefont {Tillich}}\ and\ \bibinfo {author} {\bibfnamefont {G.}~\bibnamefont {Zemor}},\ }\href@noop {} {\bibfield  {journal} {\bibinfo  {journal} {IEEE Trans. Inf. Theory}\ }\textbf {\bibinfo {volume} {60}},\ \bibinfo {pages} {1193} (\bibinfo {year} {2014})}\BibitemShut {NoStop}%
\bibitem [{\citenamefont {Roffe}\ \emph {et~al.}(2023)\citenamefont {Roffe}, \citenamefont {Cohen}, \citenamefont {Quintavalle}, \citenamefont {Chandra},\ and\ \citenamefont {Campbell}}]{Roffe2023-ay}%
  \BibitemOpen
  \bibfield  {author} {\bibinfo {author} {\bibfnamefont {J.}~\bibnamefont {Roffe}}, \bibinfo {author} {\bibfnamefont {L.~Z.}\ \bibnamefont {Cohen}}, \bibinfo {author} {\bibfnamefont {A.~O.}\ \bibnamefont {Quintavalle}}, \bibinfo {author} {\bibfnamefont {D.}~\bibnamefont {Chandra}},\ and\ \bibinfo {author} {\bibfnamefont {E.~T.}\ \bibnamefont {Campbell}},\ }\href@noop {} {\bibfield  {journal} {\bibinfo  {journal} {Quantum}\ }\textbf {\bibinfo {volume} {7}},\ \bibinfo {pages} {1005} (\bibinfo {year} {2023})}\BibitemShut {NoStop}%
\bibitem [{\citenamefont {Panteleev}\ and\ \citenamefont {Kalachev}(2022{\natexlab{a}})}]{Panteleev2022-bp}%
  \BibitemOpen
  \bibfield  {author} {\bibinfo {author} {\bibfnamefont {P.}~\bibnamefont {Panteleev}}\ and\ \bibinfo {author} {\bibfnamefont {G.}~\bibnamefont {Kalachev}},\ }in\ \href@noop {} {\emph {\bibinfo {booktitle} {Proceedings of the 54th Annual {ACM} SIGACT Symposium on Theory of Computing}}}\ (\bibinfo  {publisher} {ACM},\ \bibinfo {address} {New York, NY, USA},\ \bibinfo {year} {2022})\BibitemShut {NoStop}%
\bibitem [{\citenamefont {Weibel}(1994)}]{homological}%
  \BibitemOpen
  \bibfield  {author} {\bibinfo {author} {\bibfnamefont {C.}~\bibnamefont {Weibel}},\ }\href@noop {} {\bibfield  {journal} {\bibinfo  {journal} {An Introduction to Homological Algebra}\ ,\ \bibinfo {pages} {1–29}} (\bibinfo {year} {1994})}\BibitemShut {NoStop}%
\bibitem [{\citenamefont {Bravyi}\ and\ \citenamefont {Hastings}(2013)}]{Bravyi2013-gu}%
  \BibitemOpen
  \bibfield  {author} {\bibinfo {author} {\bibfnamefont {S.}~\bibnamefont {Bravyi}}\ and\ \bibinfo {author} {\bibfnamefont {M.~B.}\ \bibnamefont {Hastings}},\ }\href@noop {} {\bibfield  {journal} {\bibinfo  {journal} {arXiv [quant-ph]}\ } (\bibinfo {year} {2013})}\BibitemShut {NoStop}%
\bibitem [{\citenamefont {Panteleev}\ and\ \citenamefont {Kalachev}(2021)}]{Panteleev2021-oz}%
  \BibitemOpen
  \bibfield  {author} {\bibinfo {author} {\bibfnamefont {P.}~\bibnamefont {Panteleev}}\ and\ \bibinfo {author} {\bibfnamefont {G.}~\bibnamefont {Kalachev}},\ }\href@noop {} {\bibfield  {journal} {\bibinfo  {journal} {Quantum}\ }\textbf {\bibinfo {volume} {5}},\ \bibinfo {pages} {585} (\bibinfo {year} {2021})}\BibitemShut {NoStop}%
\bibitem [{\citenamefont {Xu}\ \emph {et~al.}(2023)\citenamefont {Xu}, \citenamefont {Mannucci}, \citenamefont {Seif}, \citenamefont {Kubica}, \citenamefont {Flammia},\ and\ \citenamefont {Jiang}}]{Xu2023-fw}%
  \BibitemOpen
  \bibfield  {author} {\bibinfo {author} {\bibfnamefont {Q.}~\bibnamefont {Xu}}, \bibinfo {author} {\bibfnamefont {N.}~\bibnamefont {Mannucci}}, \bibinfo {author} {\bibfnamefont {A.}~\bibnamefont {Seif}}, \bibinfo {author} {\bibfnamefont {A.}~\bibnamefont {Kubica}}, \bibinfo {author} {\bibfnamefont {S.~T.}\ \bibnamefont {Flammia}},\ and\ \bibinfo {author} {\bibfnamefont {L.}~\bibnamefont {Jiang}},\ }\href@noop {} {\bibfield  {journal} {\bibinfo  {journal} {Phys. Rev. Res.}\ }\textbf {\bibinfo {volume} {5}} (\bibinfo {year} {2023})}\BibitemShut {NoStop}%
\bibitem [{\citenamefont {Huang}\ \emph {et~al.}(2023)\citenamefont {Huang}, \citenamefont {Pesah}, \citenamefont {Chubb}, \citenamefont {Vasmer},\ and\ \citenamefont {Dua}}]{Huang2023-ni}%
  \BibitemOpen
  \bibfield  {author} {\bibinfo {author} {\bibfnamefont {E.}~\bibnamefont {Huang}}, \bibinfo {author} {\bibfnamefont {A.}~\bibnamefont {Pesah}}, \bibinfo {author} {\bibfnamefont {C.~T.}\ \bibnamefont {Chubb}}, \bibinfo {author} {\bibfnamefont {M.}~\bibnamefont {Vasmer}},\ and\ \bibinfo {author} {\bibfnamefont {A.}~\bibnamefont {Dua}},\ }\href@noop {} {\bibfield  {journal} {\bibinfo  {journal} {PRX quantum}\ }\textbf {\bibinfo {volume} {4}} (\bibinfo {year} {2023})}\BibitemShut {NoStop}%
\bibitem [{\citenamefont {Fowler}\ \emph {et~al.}(2009)\citenamefont {Fowler}, \citenamefont {Stephens},\ and\ \citenamefont {Groszkowski}}]{Fowler2009-xe}%
  \BibitemOpen
  \bibfield  {author} {\bibinfo {author} {\bibfnamefont {A.~G.}\ \bibnamefont {Fowler}}, \bibinfo {author} {\bibfnamefont {A.~M.}\ \bibnamefont {Stephens}},\ and\ \bibinfo {author} {\bibfnamefont {P.}~\bibnamefont {Groszkowski}},\ }\href@noop {} {\bibfield  {journal} {\bibinfo  {journal} {Phys. Rev. A}\ }\textbf {\bibinfo {volume} {80}} (\bibinfo {year} {2009})}\BibitemShut {NoStop}%
\bibitem [{\citenamefont {Panteleev}\ and\ \citenamefont {Kalachev}(2022{\natexlab{b}})}]{Panteleev2022-qh}%
  \BibitemOpen
  \bibfield  {author} {\bibinfo {author} {\bibfnamefont {P.}~\bibnamefont {Panteleev}}\ and\ \bibinfo {author} {\bibfnamefont {G.}~\bibnamefont {Kalachev}},\ }\href@noop {} {\bibfield  {journal} {\bibinfo  {journal} {IEEE Trans. Inf. Theory}\ }\textbf {\bibinfo {volume} {68}},\ \bibinfo {pages} {213} (\bibinfo {year} {2022}{\natexlab{b}})}\BibitemShut {NoStop}%
\bibitem [{\citenamefont {Bazarsky}\ \emph {et~al.}(2013)\citenamefont {Bazarsky}, \citenamefont {Presman},\ and\ \citenamefont {Litsyn}}]{Bazarsky2013ACE}%
  \BibitemOpen
  \bibfield  {author} {\bibinfo {author} {\bibfnamefont {A.}~\bibnamefont {Bazarsky}}, \bibinfo {author} {\bibfnamefont {N.}~\bibnamefont {Presman}},\ and\ \bibinfo {author} {\bibfnamefont {S.}~\bibnamefont {Litsyn}},\ }in\ \href@noop {} {\emph {\bibinfo {booktitle} {2013 {IEEE} Information Theory Workshop ({ITW})}}}\ (\bibinfo  {publisher} {IEEE},\ \bibinfo {year} {2013})\BibitemShut {NoStop}%
\bibitem [{\citenamefont {Bravyi}\ \emph {et~al.}(2024)\citenamefont {Bravyi}, \citenamefont {Cross}, \citenamefont {Gambetta}, \citenamefont {Maslov}, \citenamefont {Rall},\ and\ \citenamefont {Yoder}}]{Bravyi2024-am}%
  \BibitemOpen
  \bibfield  {author} {\bibinfo {author} {\bibfnamefont {S.}~\bibnamefont {Bravyi}}, \bibinfo {author} {\bibfnamefont {A.~W.}\ \bibnamefont {Cross}}, \bibinfo {author} {\bibfnamefont {J.~M.}\ \bibnamefont {Gambetta}}, \bibinfo {author} {\bibfnamefont {D.}~\bibnamefont {Maslov}}, \bibinfo {author} {\bibfnamefont {P.}~\bibnamefont {Rall}},\ and\ \bibinfo {author} {\bibfnamefont {T.~J.}\ \bibnamefont {Yoder}},\ }\href@noop {} {\bibfield  {journal} {\bibinfo  {journal} {Nature}\ }\textbf {\bibinfo {volume} {627}},\ \bibinfo {pages} {778} (\bibinfo {year} {2024})}\BibitemShut {NoStop}%
\bibitem [{\citenamefont {Akarsu}(2022)}]{Akarsu2022-vw}%
  \BibitemOpen
  \bibfield  {author} {\bibinfo {author} {\bibfnamefont {M.}~\bibnamefont {Akarsu}},\ }\href@noop {} {\bibfield  {journal} {\bibinfo  {journal} {Kuramsal E{\u g}it.}\ }\textbf {\bibinfo {volume} {15}},\ \bibinfo {pages} {64} (\bibinfo {year} {2022})}\BibitemShut {NoStop}%
\bibitem [{\citenamefont {Roffe}\ and\ \citenamefont {collaborators}(2023{\natexlab{a}})}]{ldpc_repo}%
  \BibitemOpen
  \bibfield  {author} {\bibinfo {author} {\bibfnamefont {J.}~\bibnamefont {Roffe}}\ and\ \bibinfo {author} {\bibnamefont {collaborators}},\ }\href@noop {} {\bibinfo {title} {Ldpc: Symbolic construction of classical and quantum ldpc codes}},\ \bibinfo {howpublished} {\url{https://github.com/quantumgizmos/ldpc}} (\bibinfo {year} {2023}{\natexlab{a}}),\ \bibinfo {note} {accessed May 9, 2025}\BibitemShut {NoStop}%
\bibitem [{\citenamefont {Roffe}\ and\ \citenamefont {collaborators}(2023{\natexlab{b}})}]{bposd_repo}%
  \BibitemOpen
  \bibfield  {author} {\bibinfo {author} {\bibfnamefont {J.}~\bibnamefont {Roffe}}\ and\ \bibinfo {author} {\bibnamefont {collaborators}},\ }\href@noop {} {\bibinfo {title} {Bposd: Belief propagation and ordered statistics decoding for quantum ldpc codes}},\ \bibinfo {howpublished} {\url{https://github.com/quantumgizmos/bp_osd}} (\bibinfo {year} {2023}{\natexlab{b}}),\ \bibinfo {note} {accessed May 9, 2025}\BibitemShut {NoStop}%
\bibitem [{\citenamefont {Roffe}(2023)}]{bias_tailored_repo}%
  \BibitemOpen
  \bibfield  {author} {\bibinfo {author} {\bibfnamefont {J.~e.~a.}\ \bibnamefont {Roffe}},\ }\href@noop {} {\bibinfo {title} {Bias-tailored quantum ldpc codes: Simulation framework}},\ \bibinfo {howpublished} {\url{https://github.com/quantumgizmos/bias_tailored_qldpc}} (\bibinfo {year} {2023}),\ \bibinfo {note} {accessed May 9, 2025}\BibitemShut {NoStop}%
\bibitem [{\citenamefont {Vasmer}(2023)}]{Vasmer2023-single-shot}%
  \BibitemOpen
  \bibfield  {author} {\bibinfo {author} {\bibfnamefont {M.}~\bibnamefont {Vasmer}},\ }\href {https://github.com/MikeVasmer/single_shot_3D_HGP/tree/bdfb437b2abcfa514997f26be97a711b878448cb} {\bibinfo {title} {single\_shot\_3d\_hgp}},\ \bibinfo {howpublished} {\url{https://github.com/MikeVasmer/single_shot_3D_HGP/tree/bdfb437b2abcfa514997f26be97a711b878448cb}} (\bibinfo {year} {2023}),\ \bibinfo {note} {accessed: 2025-05-14}\BibitemShut {NoStop}%
\bibitem [{\citenamefont {Richardson}\ and\ \citenamefont {Urbanke}(2001)}]{Richardson2001-trapping}%
  \BibitemOpen
  \bibfield  {author} {\bibinfo {author} {\bibfnamefont {T.}~\bibnamefont {Richardson}}\ and\ \bibinfo {author} {\bibfnamefont {R.}~\bibnamefont {Urbanke}},\ }\href@noop {} {\bibfield  {journal} {\bibinfo  {journal} {IEEE Transactions on Information Theory}\ }\textbf {\bibinfo {volume} {47}},\ \bibinfo {pages} {619} (\bibinfo {year} {2001})}\BibitemShut {NoStop}%
\bibitem [{\citenamefont {He}\ \emph {et~al.}(2018)\citenamefont {He}, \citenamefont {Zhou},\ and\ \citenamefont {Du}}]{He2018-peg}%
  \BibitemOpen
  \bibfield  {author} {\bibinfo {author} {\bibfnamefont {X.}~\bibnamefont {He}}, \bibinfo {author} {\bibfnamefont {L.}~\bibnamefont {Zhou}},\ and\ \bibinfo {author} {\bibfnamefont {J.}~\bibnamefont {Du}},\ }\href@noop {} {\bibfield  {journal} {\bibinfo  {journal} {IEEE Trans. Commun.}\ }\textbf {\bibinfo {volume} {66}},\ \bibinfo {pages} {1845} (\bibinfo {year} {2018})}\BibitemShut {NoStop}%
\bibitem [{\citenamefont {Hutter}\ \emph {et~al.}(2020)\citenamefont {Hutter}, \citenamefont {Wootton},\ and\ \citenamefont {Pryadko}}]{Pryadko2020-metabp}%
  \BibitemOpen
  \bibfield  {author} {\bibinfo {author} {\bibfnamefont {A.}~\bibnamefont {Hutter}}, \bibinfo {author} {\bibfnamefont {J.~R.}\ \bibnamefont {Wootton}},\ and\ \bibinfo {author} {\bibfnamefont {L.~P.}\ \bibnamefont {Pryadko}},\ }\href {https://doi.org/10.1088/2058-9565/ab7a5c} {\bibfield  {journal} {\bibinfo  {journal} {Quantum Science and Technology}\ }\textbf {\bibinfo {volume} {5}},\ \bibinfo {pages} {034008} (\bibinfo {year} {2020})}\BibitemShut {NoStop}%
\bibitem [{\citenamefont {Varsamopoulos}\ \emph {et~al.}(2017)\citenamefont {Varsamopoulos}, \citenamefont {Criger},\ and\ \citenamefont {Bertels}}]{Varsamopoulos_2017}%
  \BibitemOpen
  \bibfield  {author} {\bibinfo {author} {\bibfnamefont {S.}~\bibnamefont {Varsamopoulos}}, \bibinfo {author} {\bibfnamefont {B.}~\bibnamefont {Criger}},\ and\ \bibinfo {author} {\bibfnamefont {K.}~\bibnamefont {Bertels}},\ }\href {https://doi.org/10.1088/2058-9565/aa955a} {\bibfield  {journal} {\bibinfo  {journal} {Quantum Science and Technology}\ }\textbf {\bibinfo {volume} {3}},\ \bibinfo {pages} {015004} (\bibinfo {year} {2017})}\BibitemShut {NoStop}%
\bibitem [{\citenamefont {Higgott}\ and\ \citenamefont {Breuckmann}(2023)}]{Higgott2023-hz}%
  \BibitemOpen
  \bibfield  {author} {\bibinfo {author} {\bibfnamefont {O.}~\bibnamefont {Higgott}}\ and\ \bibinfo {author} {\bibfnamefont {N.~P.}\ \bibnamefont {Breuckmann}},\ }\href@noop {} {\bibfield  {journal} {\bibinfo  {journal} {PRX quantum}\ }\textbf {\bibinfo {volume} {4}} (\bibinfo {year} {2023})}\BibitemShut {NoStop}%
\bibitem [{\citenamefont {Brown}\ and\ \citenamefont {Brown}(2022)}]{Brown2022-bs}%
  \BibitemOpen
  \bibfield  {author} {\bibinfo {author} {\bibfnamefont {B.~J.}\ \bibnamefont {Brown}}\ and\ \bibinfo {author} {\bibfnamefont {K.~R.}\ \bibnamefont {Brown}},\ }\href@noop {} {\bibfield  {journal} {\bibinfo  {journal} {Phys.\ Rev.\ Appl.}\ }\textbf {\bibinfo {volume} {17}},\ \bibinfo {pages} {044082} (\bibinfo {year} {2022})}\BibitemShut {NoStop}%
\bibitem [{\citenamefont {Roberts}(2022)}]{Roberts2022-rv}%
  \BibitemOpen
  \bibfield  {author} {\bibinfo {author} {\bibfnamefont {S.~E. e.~a.}\ \bibnamefont {Roberts}},\ }\href@noop {} {\bibfield  {journal} {\bibinfo  {journal} {Quantum Sci.\ Technol.}\ }\textbf {\bibinfo {volume} {7}},\ \bibinfo {pages} {045012} (\bibinfo {year} {2022})}\BibitemShut {NoStop}%
\end{thebibliography}%

\appendix
\section{Single-Shot Decoding Algorithm}
\label{app:algo}

Algorithm~\ref{alg:decode} gives the Monte-Carlo procedure used in Sec.~\ref{sec:results}.
\vspace{-1em}

\begin{algorithm}[t]  
\caption{Single-Shot CSS Simulation with BP\,+\,OSD}
\label{alg:decode}

\KwIn{Parity–check matrices $H_X,H_Z$; metachecks $M_X,M_Z$; physical error rate $p$;
      bias vector $(\beta_X,\beta_Y,\beta_Z)$; measurement-error rate $q$;
      decoding parameters (BP method, OSD order, \dots); rounds $T$.}

\KwOut{Logical error rate (LER), word error rate (WER).}

\BlankLine
\textbf{Pre-processing}\;
\begin{enumerate}
  \item Build CSS code object from $(H_X,H_Z)$.
  \item Compute channel probabilities
        $\Pr(X)=p\!\cdot\!\beta_X/\Sigma,\;
          \Pr(Y)=p\!\cdot\!\beta_Y/\Sigma,\;
          \Pr(Z)=p\!\cdot\!\beta_Z/\Sigma$,
        where $\Sigma=\beta_X+\beta_Y+\beta_Z$.
  \item Instantiate four BP+OSD decoders:\;
        \hfill $\mathsf{BP}_X$ on $H_Z$, \quad $\mathsf{BP}_Z$ on $H_X$ (data errors)\;
        \hfill $\mathsf{BP}_{\mathrm{MX}}$ on $M_X$, \quad $\mathsf{BP}_{\mathrm{MZ}}$ on $M_Z$ (meas.\ errors).
\end{enumerate}

\BlankLine
\textbf{Main Monte-Carlo Loop} \tcp*[r]{$t = 1,\dots,T$}

\begin{enumerate}
  \item \textsc{SamplePhysicalError}\;
        Independently flip each qubit as $I,X,Y,Z$ according to channel probabilities;
        store binary vectors $e_X, e_Z$.
  \item \textsc{SyndromeGeneration}\;
        $s_X^{\mathrm{ideal}} \gets H_X e_X \bmod 2$.\;
        Add measurement noise:
        $s_{X}^{\mathrm{noisy}}\gets s_X^{\mathrm{ideal}}+\operatorname{Bernoulli}(q) \pmod 2$
        (likewise for $Z$).
  \item \textsc{MetacheckDecode}\;
        Compute meta-syndromes $m_X \gets M_X s_{X}^{\mathrm{noisy}}$,
        $m_Z \gets M_Z s_{Z}^{\mathrm{noisy}}$.\;
        Decode with $\mathsf{BP}_{\mathrm{MX}},\mathsf{BP}_{\mathrm{MZ}}$
        to obtain \emph{meas.\ error estimates}
        $\hat\mu_X,\hat\mu_Z$.
  \item \textsc{SyndromeCorrection}\;
        $\tilde s_X \gets s_{X}^{\mathrm{noisy}}+\hat\mu_X \pmod 2$,\;
        $\tilde s_Z \gets s_{Z}^{\mathrm{noisy}}+\hat\mu_Z \pmod 2$.
  \item \textsc{DataDecode}\;
        \Indp
        (a) Decode $Z$-errors first:
            $\hat e_Z \gets \mathsf{BP}_Z(\tilde s_Z)$.\;
        (b) \emph{Optional channel update:} adjust $\mathsf{BP}_X$ priors using $\hat e_Z$.\;
        (c) Decode $X$-errors:
            $\hat e_X \gets \mathsf{BP}_X(\tilde s_X)$.\;
        \Indm
  \item \textsc{FailureTest}\;
        Residual error
        $\varepsilon_X \!= e_X+\hat e_X,\;
         \varepsilon_Z \!= e_Z+\hat e_Z \pmod 2$.\;
        If $(L_Z \varepsilon_X)\lor(L_X \varepsilon_Z)\neq 0$ for any logical
        operator $L_{X/Z}$, count a \emph{logical failure}; otherwise success.
  \item Update running tallies: counts of BP convergence, OSD-0 success, OSD-W success,
        minimum logical weight.
\end{enumerate}
\end{algorithm}

\end{document}